\documentclass[twocolumn,aps,prl,10pt,amsmath,amssymb,nofootinbib,showpacs,superscriptaddress,floatfix]{revtex4-1}

\newcommand{\rs}{\rm\scriptscriptstyle}

\DeclareFontFamily{U}{rcjhbltx}{}
\DeclareFontShape{U}{rcjhbltx}{m}{n}{<->rcjhbltx}{}
\DeclareSymbolFont{hebrewletters}{U}{rcjhbltx}{m}{n}

\usepackage{graphicx}
\usepackage{color}
\usepackage[usenames,dvipsnames]{xcolor}
\usepackage[colorlinks=true,linkcolor=Red,citecolor=Green,linktoc=page]{hyperref}
\usepackage{multirow}
\usepackage{float}
\usepackage{flushend}
\usepackage{balance}
\usepackage[varg]{txfonts}
\usepackage{ulem}
\usepackage{fancyhdr}

\DeclareMathSymbol{\lamed}{\mathord}{hebrewletters}{108}

\begin{document}
\title{How planar superconductors cure their infrared divergences}

\begin{abstract}
\end{abstract}

\author{M.\,C.\,Diamantini}

\affiliation{NiPS Laboratory, INFN and Dipartimento di Fisica e Geologia, University of Perugia, via A. Pascoli, I-06100 Perugia, Italy}

\author{C.\,A.\,Trugenberger}

\affiliation{SwissScientific Technologies SA, rue du Rhone 59, CH-1204 Geneva, Switzerland}

\author{V. M. Vinokur}
\affiliation{Terra Quantum AG, St. Gallerstrasse 16A, 9400 Rorschach, Switzerland}

\
\begin{abstract}
	\noindent
Planar superconductors emerging in thin films with thickness comparable to the superconducting coherence length, differ crucially from their bulk counterparts. Coulomb interactions between charges are logarithmic up to distances comparable to typical sample sizes and the Anderson-Higgs mechanism is ineffective to screen the infrared divergences of the resulting (2+1)-dimensional QED because the Pearl length screening the vortex interactions is also typically larger than the sample size. As a result, the system decomposes into superconducting droplets with the typical size of order of superconducting coherence length. We show that two possible phases of the film match the two known mechanisms for curing the (2+1)-dimensional QED infrared divergences, either by generating a mixed topological Chern-Simons mass or by magnetic monopole instantons. The former mechanism works in superconductors, the latter one governs mirror-dual superinsulators. Planar superconductors are thus described by a topological Chern-Simons gauge (TCSG) theory that replaces Ginzburg-Landau model in two dimensions. In the TCSG model, the Higgs field is absent. Accordingly, in planar superconductors Abrikosov vortices do not form, and only Josephson vortices without normal core do exist. 

\end{abstract}
\maketitle

\noindent 
Superconducting films thinner than the superconducting coherence length $\xi$ bring into being two-dimensional (2D) superconductors whose properties differ crucially from their bulk counterparts. 
Firstly, global 2D superconductivity, where the whole system becomes phase-coherent, sets in at temperatures below the Berezinskii-Kosterlitz-Thoules (BKT) transition, $T$$\leqslant$$T_{\rs BKT}$, see\,\cite{Goldman2013,tinkham,minnhagen} for a review. In other words, `two-dimensional,' referred hereafter as `planar' superconductivity, arises via a topological phase transition. Next, planar superconductors harbor the celebrated superconductor-insulator transition (SIT),\,\cite{Goldman2010}, which has been demonstrated to be a topological phase transition as well, and phases emerging in its vicinity are identified as topological phases\,\cite{dtv2019}.

The macroscopic physics of superconductors is commonly described by the Ginzburg-Landau (GL) theory\,\cite{tinkham} neglecting usually the Coulomb interactions. 
Planar superconductivity, in films with the thickness $d\lesssim \xi$,  is still discussed within the GL framework with the order parameter having a fixed amplitude and phase\,\cite{feigelman}. The bosonic model of the SIT typically treats fluctuations of the phase in the XY model\,\cite{minnhagen} and the Coulomb interaction is still neglected, since it is modelled by the same 1/r potential as in the 3D case. Here we demonstrate that the common treatment of planar superconductivity neglects crucial aspects distinguishing films from their bulk counterparts and outline the topological gauge theory replacing the standard approaches\,\cite{topsc, higgsless}. 

We consider physical systems hosting truly 2D superconductivity and realized by thin films with $d\lesssim\xi$. This ensures that electric currents are confined exclusively within-plane and that, correspondingly, the magnetic field fluctuations are perpendicular to the plane. This, however, is not sufficient; to ensure the full 2D behavior, the electric field lines also must remain within the plane, implying that the Coulomb interaction has the 2D logarithmic potential. To that end, we consider materials experiencing the SIT. In the vicinity of the SIT, superconducting films develop a huge normal-state dielectric constant $\varepsilon$ and the Coulomb interaction maintains the 2D Coulomb logarithmic form over the spatial scale range $d$$\ll$$r$$\ll$$\Lambda_{\mathrm c}$, where the 2D screening length $\Lambda_{\mathrm c}\simeq\varepsilon d$\,\cite{BV2013} reaches macroscopic scales. For example, in the TiN films, the exemplary systems manifesting the SIT, with $d=5$\,nm, $\Lambda_{\mathrm c}\simeq 200\,\mu$m, which is analogous to the behavior of the finite-temperature Coulomb potential in 3D, with $\beta = 1/T$ playing the same role as $d$, see, e.g.,\,\cite{zarembo}.
This, in turn, implies that, on the relevant spatial scales, one can consider electric interactions as genuinely 2D. As the strong logarithmic interactions cannot be neglected, the Ginzburg-Landau model has to be coupled to the 2D electromagnetism. In the common approach, having absorbed the phase within the usual Anderson-Higgs mechanism, see e.g.,\,\cite{tinkham}, the gauge fields become massive and remain coupled to charge fluctuations of the condensate. Importantly, in thin films, the scale on which gauge fields are screened, i.e., the inverse of the Anderson-Higgs mass, is given by the Pearl length $\lambda_{\perp} = \lambda_{\rs L}^2/d$, where $\lambda_{\rs L}=\sqrt{mc^2/4\pi n_{\rs s}e^2}$ is the London screening length ($m$ and $e$ are the electron mass and charge respectively, $n_{\rs s}$ is the density of superconducting electrons, $c$ is the light speed). The Pearl length is also of a macroscopic scale and for $d\simeq\xi$, the Pearl length in the TiN films is $\approx$$100\,\mu$m, i.e., comparable to the characteristic sizes of the experimental systems.  This means that on the relevant spatial scales the gauge fields remain massless within the film. 

Let us consider the perturbative expansion around the trivial gauge vacuum corresponding to vanishing fields. The resulting model is the non-compact QED in (2+1) dimensions, which is famously infrared divergent\,\cite{polyakov, jackiw1}, since the perturbation parameter is $\propto {\rm log}\ R/a$ where $a$ is the ultraviolet (UV) cutoff, in our case the coherence length $\xi$, and $R$ the infrared (IR) cutoff, the system size. As a consequence the model becomes non-perturbative (and thus ill-defined) for system sizes $\xi \ll R \le \lambda_{\perp} $. Such superconductors epytomize what we call planar superconductors: in the formal limit $d\to 0$ every superconductor is in this class. This implies that for planar superconductors the Anderson-Higgs mechanism is ineffective. In what follows we describe how these planar superconductors cure their infrared divergences. 

There are two ways in which gauge fields in (2+1) dimensions can get rid of their divergences. The first path is to generate a mass via magnetic monopole instantons, realizing tunneling between the non-trivial topological gauge vacua\,\cite{polyakov}. The other route is to generate a topological Chern-Simons (CS) mass\,\cite{jackiw1, jackiw2}. This second way, however entails breaking the discrete ${\cal P}{\cal T}$,   parity and time-reversal invariance, symmetries. This undesired property can be circumvented, however, by the CS mass arising in a model with two gauge fields and a mixed topological term\,\cite{dst}. Depending on the magnitude of $d$, the system chooses one of these possibilities. The former way results in formation of superinsulators\,\cite{dtv1,dtv2,dtv3}, the latter track  results in planar superconductors that are the focus of present communication. Actually, there is a third possibility, which is an eliminating completely both possible condensates, Cooper pairs and vortex ones, leading to the Bose metal\,\cite{dst,das1,bm,dtv2021}. The topological CS mass comprises the product of two characteristic magnetic and electric length scales, so that the film thickness $d$ falls out. The resulting gauge screening length is not the Pearl length but the original bulk London penetration depth
\begin{equation}
\lambda_{\rm top} = {\cal O}\left( \lambda_{\rs L}\right)  \ .
\label{renlon}
\end{equation}
The topological gauge theory of superconductivity replaces the Ginzburg-Landau model for planar superconductors\,\cite{topsc, higgsless}. In this model there is no Higgs field and no Abrikosov vortices that are replaced by Josephson vortices.

Let us now describe in detail the performance of the topological gauge theory. If we freeze instantons, the system has no other choice to regularize itself but to break up the condensate into the ``perturbative" droplets of the size ${\cal O}(\xi)$. These droplets are indeed observed in thin superconducting films\,\cite{sacepe}. Each of these droplets is characterized by an independent phase of the condensate and, thus, there exist topological configurations in which the circulation of these phases over the neighboring droplets is a multiple of $2\pi$. These circulations constitute Josephson vortices see, e.g.,\,\cite{tinkham}, having a nontrivial phase structure but no normal-state core. Indeed, within the droplets, Abrikosov vortices cannot exist if the droplet size compares to $\xi$\,\cite{Likharev1979} (this is the so-called Likharev vortex explosion phenomenon). The system forms global superconductivity as the Cooper pair tunneling between the droplets establishes the global phase coherence. 

The importance of the droplet structure and vortices has already been noticed in very early studies of thin superconducting films. It has been posited that these vortices can condense\,\cite{fisher1, fisher2, fisher3, fazio, girvin} at the insulating side of the SIT. The flaw of this straightforward picture, however, is that there is no duality between ballistic Cooper pairs and Abrikosov vortices, with their normal core and resulting dissipative motion making it impossible for the Abrikosov vortices to Bose condense. Here we point out that, on the contrary, the duality is realized and condensation may be possible for Josephson vortices that can form amidst the droplets by realizing the non-trivial circulation of their local phases. These Josephson vortices have no dissipative core but only a gauge structure and are, correspondingly, ballistic. Two other crucial aspects of the existence of Josephson vortices have been remaining underexplored. First circumstance, that near enough to the SIT the parameter window in which vortices become ballistic, thus moving without dissipation, gets really wide and diverges near the transition\,\cite{ballistic1}. Second, charges and vortices are subject to topological interactions leading to generating the Aharonov-Bohm\,\cite{aharonovbohm} and Aharonov-Casher\,\cite{casher} phases. These topological interactions are the infrared-dominant ones and crucially influence the long-range properties of the system. This is ultimately the reason why the treatment of thin superconducting films only in terms of the phase governed by the XY model neglecting the most important interactions is not correct. Note that the original particle-vortex duality of the XY model (see e.g.,\,\cite{minnhagen}) is broken by the charge energy term necessary to adequately describe real superconducting films and Josephson junction arrays\,\cite{fazio}. This charge energy term can be represented as a kinetic term for the charge degrees of freedom in the action\,\cite{dst}. Adding the corresponding vortex kinetic term and the mutual statistics interactions, restores the full duality of the model.

Following\,\cite{dst,bm} we treat the condensate droplets of Cooper pairs and the vortices between them in the continuum space-time with coordinates $x = (x_0, {\bf x})$  as point-like objects by introducing the fields 
\begin{eqnarray}
Q^{\mu} &&= Q_I \int ds \ {dx_I^{\mu} \over ds} \ \delta^3 \left( x -x_I (s) \right) \ ,
\nonumber \\
M^{\mu} &&= M_J \int dt \ {dx_J^{\mu} \over dt} \ \delta^3 \left( x -x_J(t) \right) \ ,
\label{pointsources}
\end{eqnarray}
with $x_I (s)$ and $x_J (t)$ parametrizing closed (for fluctuations) or infinitely long space-time trajectories so that $\partial_\mu Q^{\mu} = 0$, $\partial_\mu M^{\mu} = 0$. We will then identify the droplet size as a scale of the necessary ultraviolet cutoff. These fields describe the integer charges and vortices that constitute the main dynamical degrees of freedom of our model. Note that both trajectories and integer charge and vortex numbers $Q_I$ and $M_J$ are fluctuating dynamical variables summed over in the partition function, as described in more detail below. The mutual statistics interactions are encoded in the Euclidean partition function by phases (we use natural units $c=1$, $\hbar =1$, $\varepsilon_0=1$) 
\begin{equation}
S_{\rm top}  =  i \ 2\pi   \int d^3x \  Q_{\mu} \epsilon_{\mu \alpha \nu} {\partial_{\alpha} \over \nabla^2} M_{\nu} \ ,
\label{effec3}
\end{equation}
where $\epsilon^{\mu \alpha \nu}$ denotes the totally antisymmetric tensor. 
Using the 3D Green function
\begin{equation}
{1\over -\nabla^2} \delta^3 (x) = {1\over 4\pi} {1\over |x|} \ ,
\label{3dG}
\end{equation} 
one obtains
\begin{eqnarray}
S_{\rm top} \left( Q_I, M_J \right) &&=  i 2\pi  Q_I M_J \Phi \left( C_I, C_J \right) \ ,
\nonumber \\
\Phi \left( C_I, C_J \right) &&= {1\over 4\pi} \int_0^1 ds \int_0^1 dt \ {dx_I^{\mu} \over ds} \epsilon_{\mu \alpha \nu}
{(x_I -x_J)^{\alpha} \over |x_I-x_J|^3} {dx_J^{\nu} \over dt} \ .
\label{gauss}
\end{eqnarray} 
For closed Euclidean trajectories $C_I$ and $C_J$, the quantity $\Phi \left( C_I, C_J \right)$ represents the integer Gauss linking number, the simplest topological knot invariant, see\,\cite{kaufmann}). If the excitations satisfy the quantization condition $Q_I M_J = {\rm integer} $ for all $I, J$, the above phase becomes trivial for closed trajectories, corresponding, e.g., to a Minkowski space-time fluctuation creating a charge and a hole that annihilate after having encircled a vortex. In general, however, Aharonov-Bohm-Casher (ABC) phases lead to non-trivial quantum interference effects that cannot be neglected.

Superficially, it looks like the topological interactions are non-local, see (\ref{effec3}), and thus difficult to treat. As was pointed out by Wilczek \cite{wilczek}, however, there exists a local formulation by coupling the charge and vortex trajectories to two gauge fields $a_{\mu}$ and $b_{\mu}$ with a mixed Chern-Simons interaction \cite{dst}, leading to the Euclidean partition function
\begin{eqnarray}
Z &&= \sum_{ \{ Q^{\mu}, M^{\mu} \} } \int {\cal D} a_{\mu} {\cal D} b_{\mu} {\rm e}^{-S} \ ,
\nonumber \\
S &&= \int d^3 x {i\over 2\pi} a_{\mu} \epsilon^{\mu \alpha \nu} \partial_{\alpha} b_{\nu} + i a_{\mu} Q^{\mu} + i b_{\mu} M^{\mu} \ .
\label{wilc}
\end{eqnarray}
This is the reason why the effective field theory of planar superconductors ${\it must}$ be formulated in terms of gauge fields.  As mentioned above, the notation in (\ref{wilc}) is a short-hand for a model that is formally defined on a lattice, with the gauge fields as real variables living on the links and the charges and vortices as integer variables defined on the sites and links. The partition function involves then Riemann integration over the real-valued gauge fields and simple summation over the integer charges and vortices. Note that, at the classical level, the equations of motion imply that the dual gauge field strengths 
\begin{eqnarray}
f^{\mu} = {1\over 2} \epsilon^{\mu \alpha \nu}  f_{\alpha \nu} = \epsilon^{\mu \alpha \nu} \partial_{\alpha} b_{\nu} =2\pi Q_{\mu} \ ,
\nonumber \\
g^{\mu} = {1\over 2} \epsilon^{\mu \alpha \nu}  g_{\alpha \nu} = \epsilon^{\mu \alpha \nu} \partial_{\alpha} a_{\nu} =2\pi M_{\mu} \ ,
\label{fields}
\end{eqnarray}
can be interpreted as the conserved charge and vortex currents respectively. We use the notation in which charges and vortices are quantized in integer multiples of $2e$ and $\pi/e = 2\pi/2e$. The charge unit $2e$ can always be written as the coupling constant in front of the effective Maxwell term, as we now discuss. Note also that we do not consider single-electron fluctuations in this paper, assuming that their gap is sufficiently large to be neglected. 

We now proceed with the usual construction of the long-distance effective field theory for this system. Once the relevant symmetry has been identified, all power-counting relevant and marginal terms consistent with this symmetry have to be added to the action. In this case the symmetry is a $U(1) \otimes U(1)$ gauge symmetry and it is thus easy to identify the possible next-order terms. They must involve two derivatives and be $U(1)$ gauge invariant. Therefore the only possible choices are simply the usual Maxwell terms for the two gauge fields,
\begin{equation}
S=\int d^3 x \  {i\over2 \pi} a_{\mu} \epsilon^{\mu \alpha \nu} \partial_{\alpha} b_{\nu} 
+{1\over 2e^2_v} f_{\mu}f_{\mu} 
+{1 \over 2e^2_q} g_{\mu} g_{\mu} 
+i a_{\mu}  Q_{\mu} +i b_{\mu}  M_{\mu} \ , 
\label{nonrelac3}
\end{equation}
The two parameters $e_{\mathrm q}^2$ and $e_{\mathrm v}^2$ define the orders of magnitude of the electric and magnetic energies of an elementary charge $2e$ having the spatial scale $d$ and an elementary flux quantum $\Phi_0=\pi/e$ possessing the spatial scale $\lambda_\perp$, 
\begin{eqnarray}
e_{\mathrm q}^2 &&= {\cal O} \left( {4e^2\over  d} \right) \ ,
\nonumber \\
e_{\mathrm v}^2 &&={\cal O} \left(  {\Phi_0^2 \over \lambda_{\perp}}\right)  = {\cal O} \left( {\pi^2 \over e^2 \lambda_{\perp} } \right) = {\cal O} \left( {\pi^2 d \over e^2 \lambda_{\rs L}^2}\right)  \ .
\label{couplings}
\end{eqnarray}
They represent the two typical energy scales in the problem, and their ratio determines the relative strength of magnetic and electric forces. Non-relativistic effects can be taken into account by considering the ``space component" $x_0 = vt$ where $v$ is the velocity of the propagation in the medium and, correspondingly, all time derivatives with respect to $x_0$ and gauge field components with index ``0" as time derivatives with the prefactor $1/v$. Of course, in this effective field theory approach we cannot determine the exact coefficients in front of the two couplings.

The two parameters $e_{\mathrm q}^2$ and $e_{\mathrm v}^2$ can be traded for the topological CS mass \cite{jackiw1, jackiw2} $m=e_{\mathrm q}e_{\mathrm v}/2\pi v$, appearing in the dispersion relation of both charges and vortices,
\begin{equation}
E = \sqrt{m^2 v^4 + v^2 p^2} \ ,
\label{dis}
\end{equation}
and one dimensionless parameter $g=e_{\mathrm v}/e_{\mathrm q} = O(d/(\alpha \lambda_L))$, where $\alpha = e^2/4\pi$ is the fine structure constant. This parameter plays the role of a dimensionless conductivity and is the quantum parameter which ``selects" between the two infrared-catastrophe-avoiding models \cite{dst, bm}. 

Given that $e_{\mathrm q}^2$ and $e_{\mathrm v}^2$ have canonical dimension [1/length] and, correspondingly, the two kinetic terms in (\ref{nonrelac3}) are infrared-irrelevant, one might be tempted to leave them out and consider anyway only the infrared-dominant mixed Chern-Simons theory. Unfortunately, this leads to wrong results and the origin of the problem lies exactly in the role of the droplet charges $Q_{\mu}$ and the Josephson vortices $M_{\mu}$. Indeed, the limit $m \to \infty$ does not commute with quantization since it involves a phase-space reduction\,\cite{csqm}. For physical applications, opposing the purely mathematical ones involving the knot theory, the topological theory has  to be {\it always} considered as the $m\to\infty$ limit of the topologically massive\,\cite{jackiw2} one, since, otherwise, the physical states are not normalizable. However, when two gauge fields are present, the ``topological limit" $e_{\mathrm q}^2 \to\infty$ and $e_{\mathrm v}^2\to\infty$ is not well defined without specifying the value of $g$ in this limit. The behavior of charges and vortices depends crucially on this value and, as a consequence, one obtains very different ground states when $g$ is varied\,\cite{dst, bm}. 

We now minimally couple the charge current $j^{\mu}$ to the real electromagnetic field $A_{\mu}$ by adding to the action the term
\begin{equation}
S \to S + {i\over 2\pi} \int d^3x \ A_{\mu} f^{\mu} \ ,
\label{mincoup}
\end{equation}
and we compute the effective action $S_{\rm eff}\left( A_{\mu}, Q_{\mu}, M_{\mu} \right)$ by integrating over the fictitious gauge fields $a_{\mu}$ and $b_{\mu}$. This gives $S_{\rm eff} = S_{\rm eff}\left( A_{\mu}, Q_{\mu} \right)$ as
\begin{eqnarray}
S_{\rm eff} &&= \int d^3x \left[ 
{e_{\mathrm q}^2 \over 2} Q_\mu { \delta_{\mu\nu}  \over -\Delta + m^2 v^2}  Q_\nu \right.\nonumber \\&& \left.+i 2 \pi  m^2v^2 Q_\mu { \epsilon_{\mu \alpha \nu} \partial_\alpha \over  \Delta (-\Delta + m^2 v^2 )} \left( M_\nu + {1\over 2\pi}  F_{\nu}\right) 
+  \right. \nonumber  \\ && \left. +  {e_{\mathrm v}^2 \over 2} \left( M_\mu + {1\over 2\pi}  F_{\mu }\right) { \delta_{\mu\nu}  \over -\Delta + m^2 v^2}  \left( M_\nu + {1\over 2\pi}  F_{\nu} \right) 
 \right]   \ ,
 \label{topact}
 \end{eqnarray}
with $ F_\mu$$=$$\epsilon_{\mu\nu\alpha} \partial_\nu A_\alpha$ being the real dual electromagnetic field strength. 
Following the typical approach in lattice gauge theories\,\cite{kogut} we retain only the self-interaction terms in\,Eq.\,(\ref{topact}), replacing  the interaction kernel with its diagonal component 
\begin{equation}
G(x-y) =  {1\over m^2v^2 - \Delta} \delta^3(x-y)  \to \ell^2 G(mv\ell) \ \delta^3(x-y) \ , 
\label{green}
\end{equation}
where the necessary UV cutoff $\ell $ represents, as anticipated, the droplet size. This is, of course, an approximation, in which the screened potential is replaced by the first term in its derivative expansion, which is the delta-function of a strength depending on the ratio of the two length scales $\ell$ and $1/mv$. We also introduce the numerical parameter $\eta = (mv\ell) G(mv\ell)$. This is a numerical constant of order ${\cal O}(1)$ depending on the dimensionless quantity $mv\ell$ and is the second parameter determining the quantum phase structure\,\cite{dst, bm}.
Equation\,(\ref{topact}) become thus
\begin{eqnarray}
S_{\rm eff} &&= \int d^3x \left[
 {1\over 2} {2\pi \ell \eta  \over g} Q^2_\mu  +  {1 \over 2}  2\pi \ell \eta g \left( M_\mu + {1\over 2 \pi}  F_\mu \right)^2  + \right.  \nonumber \\  && \left.  +
 i 2 \pi  (mv\ell)  \eta Q_\mu { \epsilon_{\mu\nu\alpha} \partial_\nu \over  \Delta }  \left(M_\alpha + {1\over 2 \pi}  F_\alpha \right)   \right]   \ .
 \label{topactdiag}
 \end{eqnarray}

Let us now show how superconductivity emerges from the global condensation of the droplet charges in a phase with no vortices, $M_{\mu}=0$. This means that droplets get connected by quantum tunnelling, forming global phase coherence (therefore no vortices), i.e. a global condensate. In this phase there are no more integer-valued charges because tunnelling percolation on the droplets forms a global condensate. Correspondingly, the original integer-valued degrees of freedom $Q_{\mu}$ become a real-valued field $H_{\mu}$ over which we have to integrate in the partition function. To that end we first solve the constraint $\partial_{\mu}H_{\mu}=0$ by  the introduction of a new gauge field $n_{\mu} $ defined as $H_{\mu } = \epsilon_{\mu \alpha \nu} \partial_\alpha n_{\nu}$. This is the gauge field since shifts of $n_{\mu}$ by derivatives leave the conserved charge currents invariant. The effective action for the electromagnetic field $A_\mu$ becomes
\begin{equation}
S_{\rm eff} = \int d^3x  \ {1 \over 2} {2\pi \ell \eta \over g} H^2_\mu  + {1\over 2} { \ell \eta g \over 2\pi} F^2_\mu
- i  (mv\ell) \eta \ n_\mu \epsilon_{\mu\alpha\nu} \partial_\alpha  A_\nu  \ ,
\label{scinter}
\end{equation}
As anticipated, this effective action is not a Ginzburg-Landau/Higgs model but, rather, a mixed Chern-Simons topological theory. The electromagnetic gauge field does not acquire its mass by the Anderson-Higgs mechanism but by the topological mass generation mechanism\,\cite{jackiw1, jackiw2} with the gauge field mass $m = e_q e_v/2\pi $. The corresponding screening length $\lambda_{\rm top}=1/mv$ is the bulk London penetration depth (\ref{renlon}). Note also that the effective coupling of the Maxwell term in the action is $O(d\ell/e^2 \lambda_{\rm top})$, as can be seen using (\ref{couplings}). This shows that the the Coulomb coupling constant is renormalized to an effective coupling $e^2 \lambda_{\rm top}/\ell$: when $\ell \approx \lambda_{\rm top}$ we have the usual electron charge, when $\ell$ becomes small the effective Coulomb interaction increases. 

In the dual phase with independent droplet phases and a vortex condensate, we set $Q_{\mu}=0$ while keeping the integration over the vortex degrees of freedom $M_{\mu}$. This phase is particularly interesting for materials characterized by very high dielectric constants, so that $v\ll1$ and we can neglect the magnetic components with respect to the electric ones in the effective action, which becomes
\begin{equation}
S_{\rm eff} \left( M_{\mu}, A_{\mu} \right) = \int d^3x \ {e_v^2  \ell^2 G \over v^2 8\pi^ 2 } \left(F_i + 2 \pi M_i  \right)^2 \ ,
\label{polac}
\end{equation}
where the latin indices ``i" denote purely spatial coordinates. This is the non-relativistic version\,\cite{monjja} of the Polyakov's compact QED\,\cite{polyakov} in which magnetic monopole instantons create the confining linear potential between the probe charges and generate the photon mass. This dual phase is a superinsulator\,\cite{dst, dtv1}.

To conclude, planar superconductors cure their infrared divergences by realizing either of two possible phases with the effective actions corresponding to the two known mechanisms of the gauge field mass generation different from the Anderson-Higgs mechanism. This Higgsless, topological superconductivty\,\cite{topsc, higgsless} is the only possibility in the 2D, but may be realized also in the 3D bulk materials\,\cite{moncon, pseudo}. Indeed, 3D bulk materials with the emergent granularity typical of planar superconductors have been recently found\,\cite{parra}. 

\smallskip

\textit{Acknowledgments--} The work of V.M.V. was supported by the Terra Quantum AG.

\end{document}